\documentclass[12pt]{iopart}
\usepackage{graphicx}
\usepackage{bm}

\newcommand{\pdiff}[2]{\frac{\partial #1}{\partial #2}}
\newcommand{\fdiff}[2]{\frac{\delta #1}{\delta #2}}
\newcommand{\ave}[1]{\left\langle {#1} \right\rangle}

\newcommand{\new}{\nonumber\\}
\newcommand{\abs}[1]{\left|{#1} \right|}
\newcommand{\ox}{\overline{x}}
\newcommand{\oy}{\overline{y}}
\newcommand{\omu}{\overline{\mu}}
\newcommand{\onu}{\overline{\nu}}

\begin{document}

\title{Effect of particle exchange on the glass transition of binary hard spheres}

\author{Harukuni Ikeda$^1$, Francesco Zamponi$^1$}
\address{
$^1$Laboratoire de physique th\'eorique, D\'epartement de physique de l'ENS, \'Ecole normale sup\'erieure, PSL University, Sorbonne Universit\'e, CNRS, 75005 Paris, France}
\ead{harukuni.ikeda@lpt.ens.fr}

\begin{abstract}
We investigate the replica theory of the liquid-glass transition for a binary mixture
of large and small additive hard spheres. We consider two different
ans\"atze for this problem: the frozen glass ansatz (FGA) in whichs
the exchange of large and small particles in a glass state is prohibited, 
and the exchange glass ansatz (EGA), in which it is allowed. 
We calculate the dynamical and thermodynamical glass
transition points with the two ans\"atze. We show that the dynamical
transition density of the FGA is lower than that of the EGA, while the
thermodynamical transition density of the FGA is higher than that of the
EGA. We discuss the algorithmic implications of these results for the density-dependence of the
relaxation time of supercooled liquids.
We particularly emphasize the
difference between the standard Monte Carlo and swap Monte
Carlo algorithms. Furthermore, we discuss the importance of 
particle exchange for estimating the configurational entropy.
\end{abstract}

\maketitle
\section{Introduction}
The relaxation time of supercooled liquids increases dramatically upon
decreasing temperature or increasing density,
and eventually exceeds the experimentally accessible time scale, 
giving rise to the glass
transition~\cite{debenedetti2001supercooled,cavagna2009supercooled}.
Despite decades of studies, the underlying mechanisms that cause
the glass transition have yet to be fully understood. One of the biggest
problems is how to define a proper order
parameter for the transition, because a typical configuration of
the glass is essentially as random as a standard liquid at slightly higher
temperatures. The replica liquid theory (RLT) considers $m$ replicas of the original system in order to
circumvent the problem~\cite{monasson1995structural,FP95}. In the liquid phase,
the $m$ replicas move independently, while in the glass phase, the $m$
replicas are confined around their center of mass and behave like a
\textit{molecule}~\cite{mezard1999thermodynamics,parisi2010mean}. Thus, one can use the correlation function of the $m$
replicas as a thermodynamic order parameter, which physically corresponds to the long-time limit of the
time correlation of a single replica in the glass phase.

The RLT was first developed for one-component
systems~\cite{mezard1999thermodynamics,parisi2005ideal} and later
extended to binary
mixtures~\cite{coluzzi1999thermodynamics,biazzo2009theory,ikeda2017decoupling}.
In the latter case, the simplest ansatz corresponds to
assuming that all replicas in a molecule are of the same
species. From the physical point of view, this assumption is tantamount to
prohibit the exchange of particles in a glass state~\cite{coluzzi1999thermodynamics,ikeda2017decoupling}, which we
hereafter refer to as the frozen glass ansatz (FGA). However, the FGA-RLT
displays unphysical behavior in the one-component
limit~\cite{coluzzi1999thermodynamics}. The entropy of the glass,
as predicted by the FGA, remains larger than that of the one-component
system by the mixing entropy, 
\begin{equation}
s_{\rm mix}=-\sum_\mu x_\mu\log x_\mu \ ,
\end{equation}
where $N_\mu$ is the number of particles of the $\mu$-th
species, $N=\sum_\mu N_\mu$, and $x_\mu=N_\mu/N$, even in the limit where all particles are identical.

\begin{figure}
\centering
\includegraphics[width=.85\textwidth]{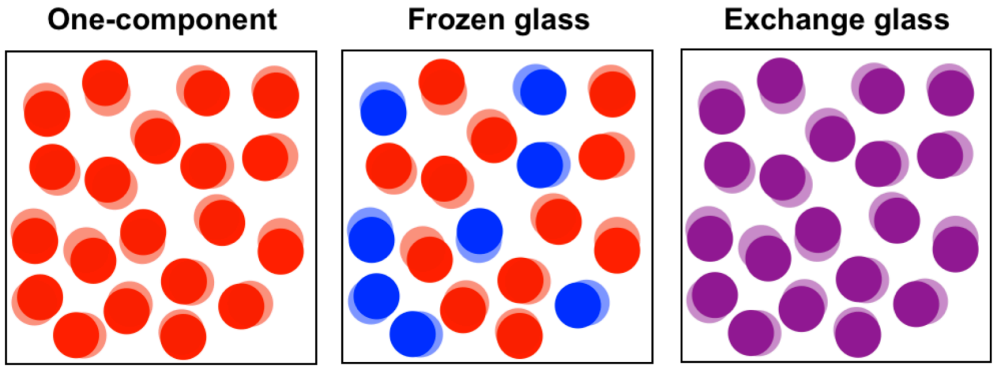}
\caption{Schematic
 examples of configurations of the $m=2$ replicated system. The number
 of the first (red) and second (blue) species are $N_1=12$ and $N_2=8$, respectively.
 The purple circles denote particles of either species. (Left) A configuration of the
 one-component system. The configuration does not change under the
 exchange of any pair of particles, and the Gibbs factor is
 $(N_1+N_2)!$. (Middle) A configuration of the FGA.  The
 configuration is only invariant under the exchange of molecules of the same species. The Gibbs factor is
 then $N_1!N_2!$. (Right) A configuration of the EGA.
 The Gibbs factor is the same as that of the one-component system,
 $G=(N_1+N_2)!$.
 }  \label{165627_15Nov18}
\end{figure}
As discussed by Coluzzi \textit{et
al.}~\cite{coluzzi1999thermodynamics}, the above discrepancy originates
from the Gibbs factor of the replicated system, i.e. the number of exchanges of
molecules that leave the configuration of the system invariant. In
Fig.~\ref{165627_15Nov18}, we show the schematic configuration of a
replicated system for $m=2$, $N_1=12$ and $N_2=8$. For one-component systems, all molecules are identical, and
the Gibbs factor is $G=(N_1+N_2)!$. For binary mixtures, if one
assumes the FGA, only the molecules of the same species can be
exchanged, leading to the Gibbs factor $G=N_1!N_2!$, 
and to an entropy difference $\Delta S= -\Delta\log G= \log (N_1+N_2)!-\log
(N_1!N_2!) = N s_{\rm mix}$ with respect to the one-component system. Note that $\Delta S$ only depends on the particle numbers, and it thus remains
constant even in the one-component limit.

To resolve this contradiction, one should allow the exchange of
replicas of different species (i.e. the dissociation of molecules), see the right panel in
Fig.~\ref{165627_15Nov18}.  In this case, the particle species in
a molecule change with time. After averaging over the time, the
molecules are indistinguishable, meaning that the Gibbs factor is
$G=(N_1+N_2)!$, and the one-component result is recovered. Hereafter, we
shall refer to this as the exchange glass ansatz (EGA). From the
physical point of view, the EGA is tantamount to allow the exchange
of particles in a glass state; the exchange process leads to the dissociation
of molecules. Note that the FGA corresponds to
the extreme case of the EGA when the probability of exchanging
particles of different species vanish.  This is indeed the case if, say,
the size ratio of different species is sufficiently large.  On the
contrary, if the size ratio is close to unity or particles have a
continuous size distribution, the EGA is needed to avoid unphysical behavior.

The difference between the FGA and EGA also sheds light into the algorithmic
dependence of the relaxation time of supercooled
liquids~\cite{ikeda2017mean,szamel2018theory} (see~\cite{wyart2017length} for an alternative approach). Recently,
it has been shown that the swap Monte Carlo (SMC) algorithm, which is nothing but a
standard Monte Carlo algorithm (MC) combined with particle exchange moves,
significantly accelerates the relaxation of supercooled
liquids~\cite{PhysRevE.63.045102,andrea2017model}. The FGA and EGA give
an effective description, within mean field theory, of the metastable states 
accessible by MC and SMC~\cite{ikeda2017mean}. Over a time
scale sufficiently shorter than the relaxation time $\tau_\alpha$, particles
undergo vibrational motion around their equilibrium position. 
The exchange of particle species hardly occurs over this time scale in
standard MC, meaning that the FGA gives a good
description of the corresponding metastable state. On the contrary, the
SMC exchanges the particle species as well as
the particle positions. Particle exchanges thus frequently occur even at
time scales shorter than $\tau_\alpha$, meaning that the corresponding metastable
state would be well described by the EGA. It is expected that the FGA
and EGA give different glass transition points, which provides an explanation for
the difference of relaxation time between the standard MC and SMC~\cite{ikeda2017mean,szamel2018theory}.

An explicit implementation of the EGA has been provided quite recently in
Ref.~\cite{ikeda2016note}. In previous work~\cite{ikeda2017mean}, we applied the formalism
of Ref.~\cite{ikeda2016note} to the Mari-Kurchan model~\cite{mari2011dynamical}, 
a mean field model of the glass
transition of binary hard spheres. In this
work, we apply the formalism to a more realistic model, i.e. binary hard
spheres in three dimensions, which allows us to calculate
quantitatively the values of dynamical and thermodynamical quantities.

The organization of the paper is as follows.  In
Section~\ref{112650_3Dec18}, we briefly summarize the mean field scenario
of the glass transition, on which the RLT relies. In
Section~\ref{112657_3Dec18}, we introduce the model.  In
Section~\ref{112707_3Dec18}, we extend the RLT to binary mixtures,
particularly emphasizing the way to take into account particle
exchange. In Sections~\ref{112741_3Dec18} and \ref{112752_3Dec18}, we
discuss the results and conclude the work.  

 \section{A mean field theory of glass transition}
 \label{112650_3Dec18} 
 
 Here, we briefly summarize the random first order
transition (RFOT) theory, which is a semi-phenomenological theory of the
glass transition built on the analogy with a class of
mean field spin-glasses~\cite{kirkpatrick1989scaling,bouchaud2004adam}.
The key assumption of RFOT theory is that the slow dynamics around the glass
transition is due to the emergence of complex structures within the free-energy
landscape. The ROFT theory predicts that the dynamics
of supercooled liquids changes qualitatively around two important points. 
The first is the dynamical transition, at packing fraction $\varphi_d$, at which
long-lived metastable states appear in the free energy landscape.  
Mean field theories of
the glass transition, such as the mode-coupling theory (MCT), predict that
the relaxation time diverges as $\tau_\alpha \sim (\varphi_d-\varphi)^{-\gamma}$~\cite{bengtzelius1984dynamics,gotze2008complex}. 
In finite dimensions, however, because of the thermal fluctuations, the
system escape from a metastable state after a sufficiently long time. The
activation events are driven by the configurational entropy $\Sigma(\varphi)$,
which is the logarithm of the number of metastable states~\cite{kirkpatrick1989scaling,bouchaud2004adam}. 
The second important point is the Kauzmann transition, at packing fraction $\varphi_K$, at
which $\Sigma(\varphi)$ vanishes. For $\varphi>\varphi_K$, the system is
permanently stuck in a free energy minimum, because it can no longer acquire entropy by
visiting several other minima. In this regime, the system is referred to as
an {\it ideal glass}, and it has a finite rigidity~\cite{yoshino2010}. Below, we
calculate the transition points $\varphi_d$ and $\varphi_K$, as well as the configurational entropy $\Sigma(\varphi)$,
of binary hard spheres by
using the RLT.

\section{Model}
\label{112657_3Dec18} 

We now introduce the model used in this work. We
consider a binary mixture of large and small particles. The total
interaction potential is
\begin{equation}
 V_N = \sum_{i<j}^{1,N} v_{\mu_i\mu_j}(x_i-x_j) \ ,
\end{equation}
where $x_i$ denotes the position of the $i$-th particle, and $\mu_i\in
\{\rm Large,\ Small\}$ denotes the particle species. We use the additive hard
sphere potential:
\begin{equation}
 v_{\mu\nu}(r) =
  \left\{
  \begin{array}{ll}
   \infty& r < \frac{\sigma_\mu+\sigma_\nu}{2}\\
   0  &  r \ge \frac{\sigma_\mu+\sigma_\nu}{2}
  \end{array}
  \right.,\label{172942_30Nov18}
\end{equation}
where $\sigma_{\rm Large}$ and $\sigma_{\rm Small}$ denote the radius of
large and small particles, respectively.
The total number of particles is $N$, and the volume of the
system is $V$.  For hard spheres, the temperature $T$ does not affect
thermodynamic quantities~\cite{hansen1990theory}. The relevant
control parameters are the packing fraction $\varphi=
V^{-1}\sum_{i=1}^N\frac{4\pi\sigma_{\mu_i}^3}{3}$, the size ratio
$r=\sigma_{\rm Large}/\sigma_{\rm Small}$, and the fraction of the
species $x_{\rm Large} = N_{\rm Large}/N$ or $x_{\rm Small}=1-x_{\rm
Large}$.

\section{Derivation of the RLT equations}
\label{112707_3Dec18}

Here, we derive the RLT equations for binary hard spheres. For this purpose, we
extend the quantitative approximation scheme that has been developed in~\cite{mangeat2016quantitative}
for one-component hard spheres.

\subsection{Effective potential method}

Our starting point is the replicated partition
function~\cite{mezard1999thermodynamics,parisi2010mean,ikeda2016note}
\begin{equation}
 Z_m = \sum_{N=0}^\infty \frac{1}{N!} \left(\prod_{a=1}^m
	\prod_{i=1}^N \sum_{\mu^a_i}\int dx_i^a\right)
  e^{-\beta\sum_{a=1}^m
	\sum_{i<j}^{1,N}v_{\mu_i\mu_j}(x_i^a-x_j^a)
	+ \sum_{i=1}^N \psi_{\omu_i}(\ox_i) },
\end{equation}
where $\ox_i=\{x_i^1,\cdots,x_i^m\}$ and
$\omu_i=\{\mu_i^1,\cdots,\mu_i^m\}$ denote the position and species of
the $i$-th particle in the replica space.  $\psi$ is the chemical
potential conjugated to the one point replicated density distribution,
\begin{equation}
 \rho_{\omu}(\ox) = \sum_{i=1}^N \ave{\prod_{a=1}^m\delta(x^a-x_i^a)\delta_{\mu^a\mu_i^a}}
  = \fdiff{\log Z_m}{\psi_{\omu}(\ox)} \ ,
\end{equation}
which is directly related to the order parameter.
For instance, the cage size $A$ is given by 
\begin{eqnarray}
 A &= \sum_{\omu}\int d\ox \rho_{\omu}(\ox)(x^a-x^b)^2 \ ,
\end{eqnarray}
for an arbitrary pair $ab$.
In order to determine $\rho_{\omu}(\ox)$, one first performs a Legendre
transformation, which expresses the free energy as a functional of
$\rho_{\omu}(\ox)$, and then optimizes the transformed free energy w.r.t $\rho_{\omu}(\ox)$. 
For the calculation of the replicated free energy, we
use the effective potential method, which allows us to map the
replicated system onto a non-replicated system, by integrating out the
degree of freedom of $m-1$ replicas~\cite{parisi2010mean}. Let us
choose the first replica $a=1$ as a reference. The effective potential of the
first replica is
\begin{eqnarray}
 e^{-\beta v^{\rm eff}_{\mu^1\nu^1}(x^1-y^1)} &= \frac{1}{x_{\mu_1}x_{\nu_1}}
  e^{-\beta v_{\mu^1\nu^1}(x^1-y^1)}
\left(  \prod_{a=2}^m \sum_{\mu^a\nu^a}  \int dx^a dy^a \right) \times \new
&\times \rho_{\omu}(\ox)\rho_{\onu}(\oy)
  e^{-\beta \sum_{a=2}^m v_{\mu^a\nu^a}(x^a-y^a)}\new
  &= e^{-m\beta v_{\mu^1\nu^1}(x^1-y^1)}\left(1 + Q_{\mu^1\nu^1}(x^1-y^1)\right),\label{151511_18Nov18}
\end{eqnarray}
where
\begin{eqnarray}
 Q_{\mu^1\nu^1}(x^1-y^1) &= -1 + \frac{e^{-(1-m)\beta v_{\mu^1\nu^1}(x^1-y^1)}}{x_\mu^1 x_\nu^1} \times \new
&\times  \left(  \prod_{a=2}^m \sum_{\mu^a\nu^a}  \int dx^a dy^a \right)  \rho_{\omu}(\ox)\rho_{\onu}(\oy)
  e^{-\beta \sum_{a=2}^m v_{\mu^a\nu^a}(x^a-y^a)} \ .
  \label{120321_26Nov18}
\end{eqnarray}
We approximate the replicated free energy by the free energy of the
first replica, interacting with the effective potential defined in
Eq.~(\ref{151511_18Nov18}),
\begin{eqnarray}
 -\beta F_m = \log Z_m  \approx \Delta S_{\rm vib} -m\beta F[v^{\rm eff}] \ , \label{153825_18Nov18}
\end{eqnarray}
where $\Delta S_{\rm vib}$ is the additional vibrational entropy coming from the other $m-1$ replicas in the ideal gas term,
\begin{equation}
 \Delta S_{\rm vib} = \sum_{\omu}\int d\ox \rho_{\omu}(\ox)(1-\log\rho_{\omu}(\ox))
  - N\sum_\mu x_\mu(1-\log\rho_\mu) \ . \label{123612_26Nov18}
\end{equation}
In a glass state, the $m$ replicas undergo vibrational motion
around their center of mass, suggesting that $Q_{\mu^1\nu^1}$ can be considered as
a small perturbation.  Expanding Eq.~(\ref{153825_18Nov18}) for small $Q_{\mu^1\nu^1}$,
we obtain
\begin{eqnarray}
-\beta F_m &\approx \Delta S_{\rm vib} -m\beta F(T^*) +
 \frac{1}{2}\sum_{\mu\nu}\rho_{\mu}\rho_{\nu}
 \int dx dy g_{\mu\nu}(x,y)Q_{\mu\nu}(x-y) \ , \label{121251_26Nov18}
\end{eqnarray}
where $\rho_\mu = x_\mu \rho$, $\rho = N/V$, $F(T^*)$ denotes the free
energy of the non-replicated liquid with the bare potential, at the effective temperature $T^*=T/m$, and 
$g_{\mu\nu}(x,y)$ denotes its
pair correlation function. Finally, we replace
the bare potential in $Q_{\mu\nu}$ as $e^{-\beta v_{\mu\nu}(r)}\to
g_{\mu\nu}(r)^{1/m}$, which has been shown to significantly improve the
accuracy of the approximation in the one-component
system case~\cite{mangeat2016quantitative}.

Full optimization of the free energy for completely general
$\rho_{\omu}(\ox)$ is very difficult. To simplify this calculation, we
approximate $\rho_{\omu}(\ox)$ as
\begin{eqnarray}
 \rho_{\omu}(\ox) \approx \rho(\ox)g(\omu) \ , \label{150150_12Oct18}
\end{eqnarray}
where 
\begin{equation}
\rho(\ox)=\sum_i\ave{\prod_a\delta(x^a-x_i^a)} \ , \qquad
g(\omu)=\frac1N \sum_i\ave{\prod_a\delta_{\mu^a\mu_i^a}} . 
\end{equation}
In principle, this
assumption can be relaxed, but the calculation becomes very
involved~\cite{ikeda2017mean}. For the
positional degrees of freedom, we assume the standard Gaussian
ansatz~\cite{mezard1999thermodynamics,parisi2010mean}:
\begin{equation}
 \rho(\ox) = \rho \int dX \prod_{a=1}^m \gamma_A(x^a-X) \ , \qquad \gamma_A(r)=\frac{e^{-r^2/2A}}{(2\pi A)^{d/2}} \ ,
 \label{151856_12Oct18}
\end{equation}
which implies that the cage
vibrations of the $m$ replicas follow the Gaussian distribution around
their center of mass $X$. The remaining task, which we shall discuss in the following subsections, is to formulate an ansatz
for $g(\omu)$.

  \subsection{Frozen Glass Ansatz (FGA)}

If the size ratio $r$ is sufficiently large, particles of different species
cannot be exchanged in a glass state. This frozen situation is described by a simple
ansatz, in which each molecule of the replicated system
consists of particles of the same species, as illustrated in the middle
panel in Fig.~\ref{165627_15Nov18}. More concretely, we assume
\begin{equation}
 g(\omu) = \sum_{\mu}x_{\mu}\prod_{a=1}^m\delta_{\mu\mu^a}.\label{151928_12Oct18}
\end{equation}
Substituting this into the vibrational entropy,
Eq.~(\ref{123612_26Nov18}), one obtains
\begin{equation}
 \frac{\Delta S_{\rm vib}}N = \frac{d}{2}(m-1)[1+\log(2\pi A)]
  + \frac{d}{2}\log m - (m-1)\sum_\mu x_\mu \log x_\mu \ .
\end{equation}
Also, Eq.~(\ref{120321_26Nov18}) reduces to
\begin{eqnarray}
 Q_{\mu\nu}(r) &= -1 + g_{\mu\nu}(r)^{(1-m)/m} 
  \int dr' \gamma_{2A}(r+r')q_{\mu\nu}(r')^{m-1},\new
  q_{\mu\nu}(r) &= \int dr' \gamma_{2A}(r+r')g_{\mu\nu}(r)^{1/m} \ .\label{121243_26Nov18}
\end{eqnarray}
The equilibrium value of the cage size $A$ is obtained from the saddle
point condition $\left.\partial_A\log Z_m\right|_{m=1}=0$. Substituting
Eq.~(\ref{121243_26Nov18}) into Eq.~(\ref{121251_26Nov18}), one gets
\begin{equation}
 A = M(A), 
\end{equation}
where
\begin{equation}
 M(A) = \left[-\frac{\rho}{d}\sum_{\mu\nu}x_\mu x_\nu
	 \int dr \pdiff{q_{\mu\nu}(r)}{A}\log q_{\mu\nu}(r)\right]^{-1}.
\end{equation}
The configurational entropy $\Sigma$ is calculated by using the
Monasson's formula~\cite{monasson1995structural}:
\begin{equation}
 \Sigma = \lim_{m\to 1}m^2 \pdiff{}{m}\left(\frac{\beta F_m/N}{m}\right).
  \label{170349_26Nov18}
\end{equation}
Using above equation, it is straightforward to show that
\begin{eqnarray}
 \Sigma &= s_{\rm liq} -\frac{d}{2}\log(2\pi A) -d \new &
  -\frac{\rho}{2}\sum_{\mu\nu}x_\mu x_\nu \int dr
  \left[q_{\mu\nu}(r)\log q_{\mu\nu}(r)-g_{\mu\nu}(r)\log g_{\mu\nu}(r)\right] \ ,\label{110331_19Oct18}
\end{eqnarray}
where $s_{\rm liq}$ is the entropy per particle of the bare liquid at
temperature $T^*$.  In the one-component limit, $r\to 1$, the FGA gives
a pathological result.  In this limit, the interaction term converges to
the one-component result $-\frac{\rho}{2}\int dr \left[q(r)\log
q(r)-g(r)\log (r)\right]$, while the liquid entropy becomes $s_{\rm liq}
= s_{\rm liq}^{\rm one} + s_{\rm mix}$.  Thus, the difference between
$\Sigma$ of the one-component and two-component systems does not vanish,
$\Delta \Sigma \to s_{\rm mix} > 0$. This unphysical behavior implies
that the FGA must break down around $r\sim 1$. In the next section, we
show that this problem can be cured by taking into account particle
exchange.

\subsection{Exchange Glass Ansatz (EGA)}

\begin{figure}[t]
\centering
 \includegraphics[width=15cm]{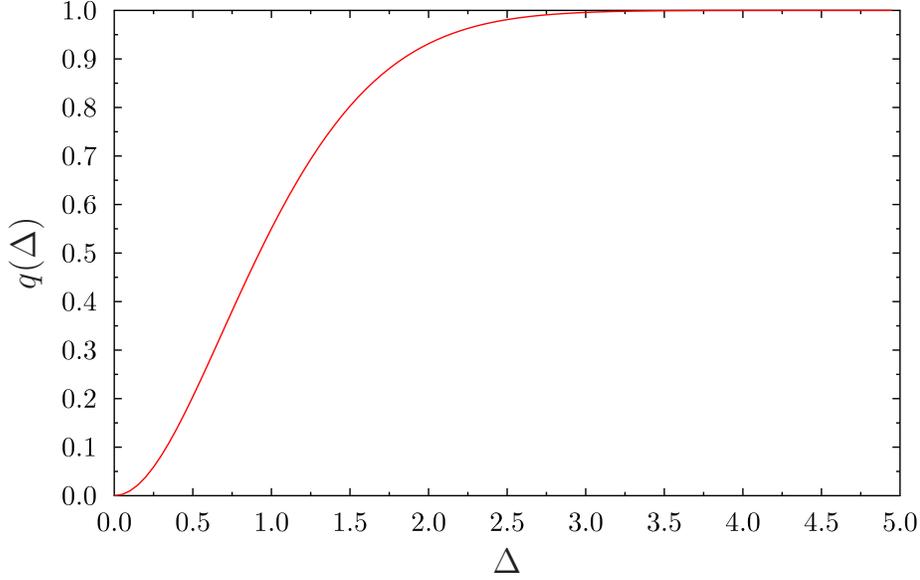}
 \caption{Dependence of the order parameter for particle exchange, $q(\Delta)$, on the coupling $\Delta$.} \label{152610_15Oct18}
\end{figure}

Here we derive the free energy within an ansatz
that takes into account the effect of particle exchange in a glass state. 
For this purpose, we first map the
particle species to a binary spin variable by introducing
$\sigma(\mu)$, where $\sigma(\rm Large)=1$ and $\sigma(\rm Small)=-1$. We
assume that $\sigma(\mu)$ follows the same distribution as 
a mean field spin glass model~\cite{mezard1987spin,ikeda2017mean},
\begin{eqnarray}
 g(\omu) &= C_m(\Delta)^{-1} e^{H\sum_{a=1}^m \sigma(\mu^a)+
  \frac{\Delta^2}{2}\sum_{ab}^{1,m}\sigma(\mu^a)\sigma(\mu^b)}\new
  &= C_m(\Delta)^{-1} \int Dh e^{(h+H)\sum_{a=1}^m \sigma(\mu^a)},\label{113629_13Oct18}
\end{eqnarray}
where 
\begin{equation}\label{eq:Dhdef}
Dh = dh \frac{1}{\sqrt{2\pi \Delta^2}}e^{-\frac{h^2}{2\Delta^2}} \ ,
\end{equation}
and $C_m(\Delta) = \int Dh \left(\sum_{\mu}
e^{(h+H)\sigma(\mu)}\right)^m$ is the normalization constant.  The value
of $H$ fixes the species concentration $x_\mu$, which is calculated as $x_\mu
= \sum_{\omu}g(\omu)\delta(\mu^a,\mu)$. To simplify the
calculation, hereafter we consider the equimolal binary mixture with $x_L=x_S
= 1/2$, which corresponds to $H=0$. The value of $\Delta$
controls the correlation of the particle species among different
replicas,
\begin{eqnarray}
 q(\Delta) &\equiv \lim_{m\to 1}\frac{2}{m(m-1)}\sum_{a<b}\ave{\sigma(\mu^a)\sigma(\mu^b)}\new
  &= \lim_{m\to 1}\sum_{\omu}\int d\ox \rho_{\omu}(\ox)\sigma(\mu^1)\sigma(\mu^2)\new
  &= C_1(\Delta)^{-1} \int Dh 2\cosh(h)\tanh(h)^2.\label{130702_28Nov18}
\end{eqnarray}
In Fig.~\ref{152610_15Oct18}, we show that $q(\Delta)$ is a monotonically
increasing function of $\Delta$.
The value $q=0$ corresponds to completely uncorrelated species in a molecule, while
the value $q=1$ corresponds to molecules of identical species, i.e. to the FGA.

Substituting Eq.~(\ref{113629_13Oct18}) into Eq.~(\ref{123612_26Nov18}),
one can calculate the vibrational entropy as
\begin{eqnarray}
\frac{ \Delta S_{\rm vib}}{N} &= \frac{d}{2}(m-1)\log(2\pi A) + \frac{d}{2}(m-1+\log m) + \log C_m(\Delta)\new
  &-\frac{\Delta^2}{2}(m+m(m-1)q_m(\Delta)) + \sum_\mu x_\mu\log x_\mu.\label{174135_26Nov18}
\end{eqnarray}
Similarly, Eq.~(\ref{120321_26Nov18}) reduces to 
\begin{eqnarray}
 Q_{\mu\nu}(r) &= -1 + g_{\mu\nu}(r)^{(1-m)/m}
  \frac{1}{x_{\mu}x_{\nu}}
  \int \frac{Du Dv}{C_m^2}e^{u\sigma(\mu)+ v\sigma(\nu)}\new
  &\times \int dr' \gamma_{2A}(r+r')q(r,u,v)^{m-1} \ ,
  \label{174143_26Nov18}
\end{eqnarray}
where $Du, Dv$ are defined in Eq.~(\ref{eq:Dhdef}) and
\begin{equation}
 q(r,u,v) = \sum_{\mu\nu}e^{u\sigma(\mu)+v\sigma(\nu)}\int dr' \gamma_{2A}(r+r')g_{\mu\nu}(r')^{1/m} \ .
\end{equation}
Substituting Eqs.~(\ref{174135_26Nov18}) and (\ref{174143_26Nov18}) into
Eq.~(\ref{121251_26Nov18}), we obtain the replicated free energy.  One
can derive the self-consistent equations for $A$ and $\Delta$ from the
extremization condition $\partial_A \log Z_m =\partial_\Delta \log Z_m =
0$. Because the derivation is
straightforward, but the result is cumbersome, we do not show the explicit expressions here. Using the
Monasson's formula, Eq.~(\ref{170349_26Nov18}), we obtain the
configurational entropy as
\begin{eqnarray}
 \Sigma &= s_{\rm liq}-\frac{d}{2}\log(2\pi A) -d
  -\frac{f(\Delta)}{2} + \frac{\Delta^2}{2}(1+q(\Delta))\new
  &-\frac{\rho}{2}\int dr \Bigg{\{}\frac{e^{-\Delta^2}}{4}
  \int Du Dv q(r,u,v)\left[\log q(r,u,v)-f(\Delta)\right]\new
  &-\sum_{\mu\nu}x_\mu x_\nu g_{\mu\nu}(r)\log g_{\mu\nu}(r)
			  \Bigg{\}} \ ,
			  \label{161744_15Oct18}
\end{eqnarray}
where we have introduced an auxiliary function:
\begin{equation}
 f(\Delta) \equiv \lim_{m\to 1}\frac{2}{C_m(\Delta)}\pdiff{C_m(\Delta)}{m}
  = e^{-\frac{\Delta^2}{2}}\int Dh \, 2\cosh(h)\log(2\cosh(h)) \ .
\end{equation}
In the one-component limit, $r\to 1$, one can show that $\Delta\to 0$
and $-f(\Delta)/2 \to -\log(2)= -s_{\rm mix}$, which exactly cancels out
the mixing entropy in $s_{\rm liq}$.  Thus, we recover the
configurational entropy of the one-component system obtained in previous
work~\cite{mangeat2016quantitative}:
\begin{equation}
\Sigma \to s_{\rm liq}^{\rm one} -\frac{d}{2}\log(2\pi A) -d -\frac{\rho}{2}\int dr
 \left[q(r)\log q(r)-g(r)\log g(r)\right].
\end{equation}


\subsection{Numerics}
Here we summarize how to calculate $A$ and $\Delta$ numerically for a
given $\varphi$ and $r$. Our theory requires the liquid entropy per particle $s_{\rm liq}$
and the pair correlation function $g_{\mu\nu}(r)$ as input. 
Following previous work~\cite{biazzo2009theory,mangeat2016quantitative}, we use a
binary version of the Carnahan-Starling (CS)
approximation~\cite{santos2005equation} for $s_{\rm liq}$, and the
Verlet-Weis approximation~\cite{verlet1972perturbation} for
$g_{\mu\nu}(r)$, see~\ref{130420_28Nov18} for details. To solve the
self-consistent equations for $A$ and $\Delta$, we first set $A=A_{\rm
ini}$ and $\Delta=\Delta_{\rm ini}$ for the initial conditions, where we
selected $A_{\rm ini}=10^{-3}$ and $\Delta_{\rm ini}=10^{-1}$ (and we
confirmed that the final results are independent of this choice if
$A_{\rm ini}$ and $\Delta_{\rm ini}$ are sufficiently small). Then, we
solve the self-consistent equations using a standard iterative
method. When $\varphi<\varphi_d$, this iterative process does not
converge, which indicates that $A\to \infty$. In practice, we stop the
calculation when $A$ exceeds unity during the iteration process. When
$\varphi>\varphi_d$, $A$ and $\Delta$ converge to finite values. In this
case, we stop the calculation when $\abs{(A_{i+1}-A_i)/A_i}<10^{-10}$.
By substituting $\Delta$ into Eq.~(\ref{130702_28Nov18}), one gets the
correlation $q(\Delta)$ of particle species.

\section{Results}
\label{112741_3Dec18}
In this section, we present numerical results for the order
parameters, $A$ and $q(\Delta)$, and the configurational entropy $\Sigma$.

 \subsection{Order parameter}
\begin{figure}[t]
\centering \includegraphics[width=15cm]{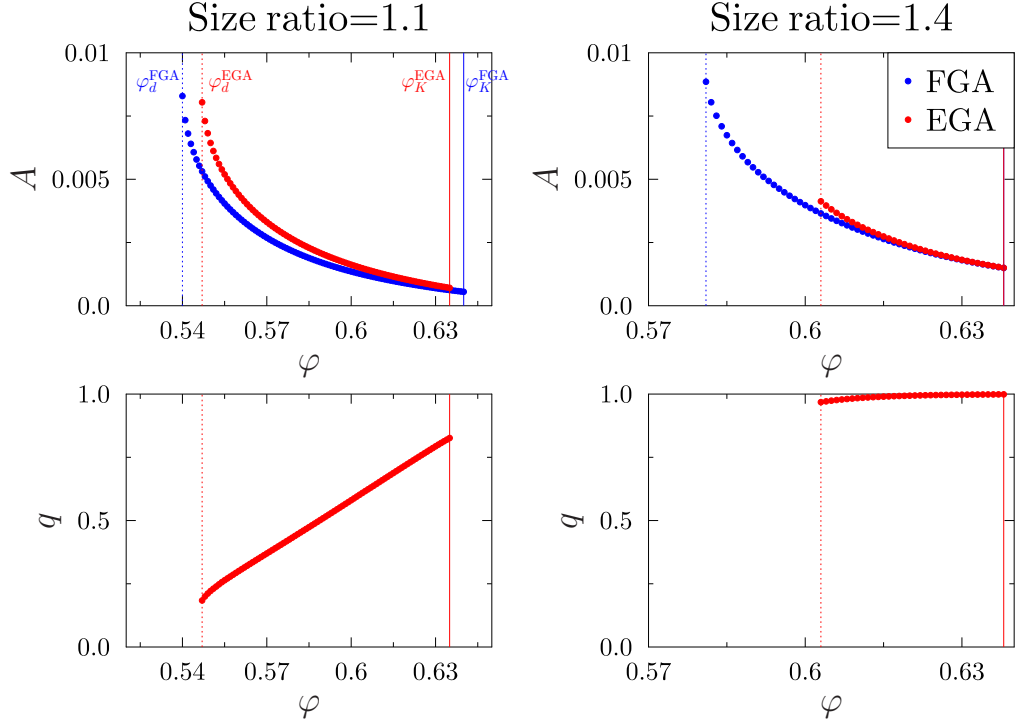}
 \caption{
 Dependence of the order parameters on packing fration $\varphi$. (Top) The cage
 size $A$. The results of the FGA and EGA are shown with blue and red
 markers, respectively. The dashed and full vertical lines indicate the
 dynamical and Kauzmann transition points, respectively.  (Bottom) The
 correlation of particle species $q$ calculated with the
 EGA. 
 } \label{185008_17Oct18}
\end{figure}
 
In Fig.~\ref{185008_17Oct18}, we show the numerical results for $A$ and
$q(\Delta)$ for size ratios $r=1.1$ and $r=1.4$.
For $r=1.1$ (left panels in
Fig.~\ref{185008_17Oct18}), for sufficiently small $\varphi$, $A=\infty$
and $q=0$, meaning that there is no correlation between different
replicas, and the system behaves as a standard liquid. $A$ and $q$ begin
to have finite values at packing fraction $\varphi=\varphi_d$,
corresponding to the dynamical transition. The dynamical
transition point of the FGA is smaller than that of the EGA,
$\varphi_d^{\rm FGA}<\varphi_d^{\rm EGA}$. 
As mentioned in the introduction, this might explain the efficiency of
the SMC reported in recent numerical simulations~\cite{andrea2017model},
if one identifies $\varphi_d^{\rm FGA}$ and $\varphi_d^{\rm EGA}$ as the
dynamical transition points of the standard MC and SMC,
respectively~\cite{ikeda2017mean}. For the same $\varphi$, the value of $A$ calculated with the FGA is
smaller than that of the EGA, which is also consistent with
numerical results~\cite{andrea2017model}.  Above 
$\varphi_d^{\rm EGA}$, $q$ increases with $\varphi$. This is a natural result, because
particle exchanges hardly occur at high $\varphi$. For very high
$\varphi$, the effect of particle exchange is negligible, and the values of $A$
calculated with the FGA and EGA are similar.  The difference
between $\varphi_d^{\rm FGA}$ and $\varphi_d^{\rm EGA}$ 
increases
with increasing $r$ (right panels in
Fig.~\ref{185008_17Oct18}). Above $\varphi_d^{\rm EGA}$, the FGA and EGA
give very similar values of $A$, which is consistent with the higher
value of $q$ above $\varphi_d^{\rm EGA}$.

\subsection{Configurational entropy}
\label{sec:sigma}

We calculate the configurational entropy $\Sigma$ by substituting the
order parameters calculated in the previous section into
Eqs.~(\ref{110331_19Oct18}) and (\ref{161744_15Oct18}). $\Sigma$ is well
defined only for $\varphi>\varphi_d$, because otherwise the order
parameters do not have finite values, indicating that there are no
metastable states.
\begin{figure}
\centering \includegraphics[width=15cm]{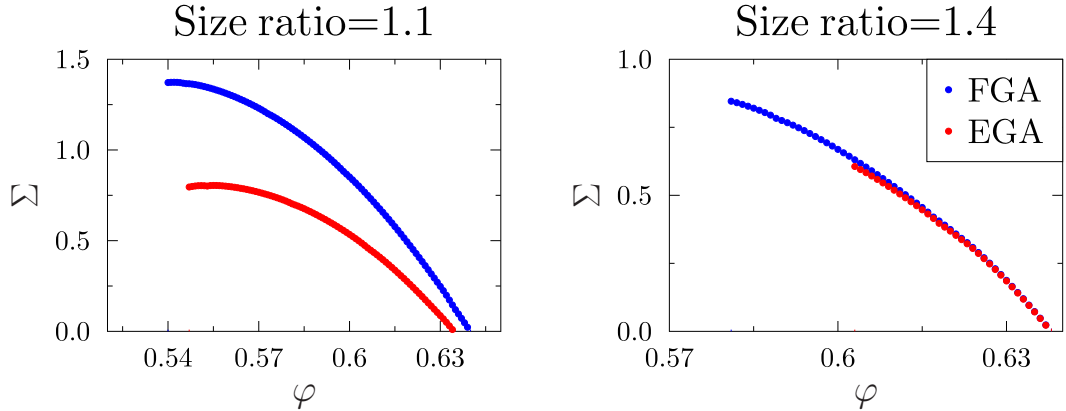} 
\caption{
Dependence of the configurational entropy $\Sigma$ on packing fraction $\varphi$,
for size ratio $r=1.1$ (left) and $r=1.4$ (right). The blue and red markers indicate the results
 of the FGA and EGA, respectively. }
						  \label{112835_19Oct18}
\end{figure}
In Fig.~\ref{112835_19Oct18}, we show the numerical results for
small and large size ratios, $r=1.1$ and $r=1.4$. For $r=1.1$, the complexity
calculated with the FGA has a higher value than that calculated with
the EGA. This results can be naturally understood as follows. The complexity 
is the difference between the entropies of the liquid and the
(metastable) glass, $\Sigma= s_{\rm liquid}-s_{\rm glass}$. In general,
$s_{\rm glass}$ calculated with the EGA is larger than with the FGA,
since the EGA includes extra degree of freedom (the species) in the glass description. As a
consequence, the EGA predicts a lower value of $\Sigma$.  
The Kauzmann transition point, $\varphi_K$, is defined by
$\Sigma(\varphi_K)=0$.  Due to the higher value of $\Sigma$, the FGA
predicts a higher value of the Kauzmann transition with respect to
the EGA, $\varphi_K^{\rm FGA}>\varphi_K^{\rm EGA}$. A similar trend is
observed for the larger size ratio $r=1.4$, see the right panel in
Fig.~\ref{112835_19Oct18}.  The difference of $\Sigma$ calculated with
the FGA and EGA decreases with increasing $r$, simply because
large and small particle are hardly exchanged.

\subsection{Phase diagram}
\begin{figure}
 \centering \includegraphics[width=13cm]{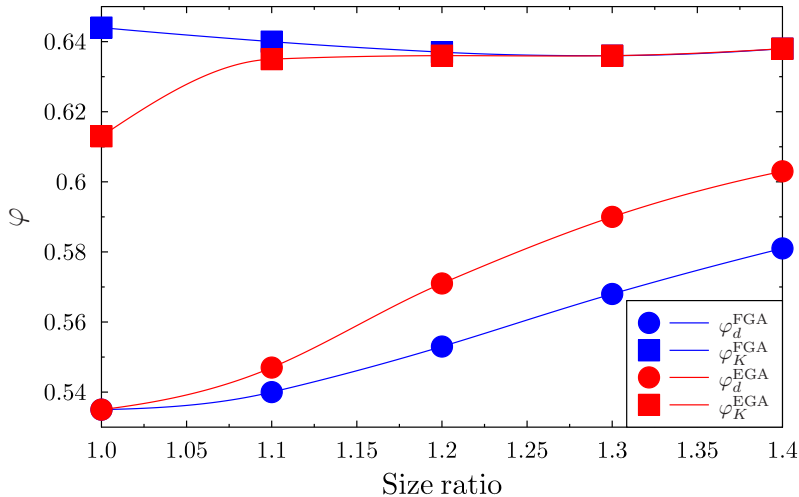} \caption{Phase
 diagram of equimolar binary hard spheres in three dimensions. Blue
 and red symbols represent the results of the FGA and EGA, respectively.
 Circles represent the dynamical transition point $\varphi_d$,
 while squares represent the Kauzmann transition
 $\varphi_K$. The solid lines are guides to the eye.}
 \label{133823_19Oct18}
\end{figure}

In Fig.~\ref{133823_19Oct18}, we show the phase diagram predicted by our
theory.
We first discuss the dynamical transition points $\varphi_d^{\rm FGA}$ and $\varphi_d^{\rm EGA}$. 
For $r>1$, we
always obtain $\varphi_d^{\rm FGA}<\varphi_d^{\rm EGA}$. This is
consistent with recent results from mode coupling theory (MCT)~\cite{szamel2018theory}. However, MCT predicts
that $\varphi_d^{\rm FGA}$ is almost independent of $r$, while our
theory predicts that $\varphi_d^{\rm FGA}$ monotonically increases with
$r$. This discrepancy may come from the approximation made in
Eq.~(\ref{150150_12Oct18}), where we assumed that the cage size $A$ does
not depend on the species. In principle one can avoid this approximation
and construct a more accurate theory, but the calculation gets much
harder~\cite{ikeda2017mean}. We leave this investigation for future work.

The behavior of the Kauzmann transition points, $\varphi_K^{\rm FGA}$ and
$\varphi_K^{\rm EGA}$, is qualitatively different from that of
$\varphi_d^{\rm FGA}$ and $\varphi_d^{\rm EGA}$. We observe that
$\varphi_K^{\rm FGA} > \varphi_K^{\rm EGA}$ for all $r$, which is a
consequence of the behavior of the configurational entropy $\Sigma$
described in Sec.\ref{sec:sigma}. 
The difference between $\varphi_K^{\rm FGA}$ and $\varphi_K^{\rm EGA}$ decreases with increasing $r$. This is a
natural result because large and small particle are hardly exchanged for
large $r$, and in particular at high density near $\varphi_K$. We
would like to stress that the FGA is metastable with respect to the EGA,
because the EGA is a more general ansatz within a variational theory. This
means that the thermodynamically meaningful transition point is not
$\varphi_K^{\rm FGA}$ but $\varphi_K^{\rm EGA}$. 
One should then take into account the degrees of freedom associated to particle
exchange when calculating the entropy of the glass state, otherwise 
the Kauzmann transition point is overestimated. A numerical algorithm for this
purpose has been recently proposed in~\cite{ozawa2018configurational}.

\section{Summary and discussions}
\label{112752_3Dec18} 

In this work, we theoretically investigated a
binary hard sphere mixture by using the replica liquid theory (RLT).  For this
purpose, we constructed two different ans\"atze: the frozen
glass ansatz (FGA), where the replicas in similar position are constrained to be of
the same species, and the exchange glass ansatz (EGA), where the replicas
in similar position can have different species. Using these ans\"atze,
we calculated the transition points for different size ratio $r$.  We
found that the dynamical transition point calculated using the FGA is
smaller than that of the EGA, $\varphi_d^{\rm FGA}<\varphi_d^{\rm EGA}$. 
The opposite relation holds for the Kauzmann transition
point, $\varphi_K^{\rm FGA}>\varphi_K^{\rm EGA}$. In the rest of this section,
we discuss possible implications of our results for experimental
and numerical studies of the glass transition.

As mentioned in the introduction, our theoretical results might give 
insight on the increased efficiency of the swap Monte Carlo algorithm
(SMC), as compared 
to the standard Monte Carlo algorithm (MC)~\cite{andrea2017model,berthier2017configurational}. 
For this discussion, it is useful to introduce two timescales: the
density relaxation time, $\tau_\alpha$, and the typical timescale to exchange
large and small particles in a metastable glass state, $\tau_{\rm ex}$. 
Over a time scale sufficiently shorter than the relaxation time, $t\ll\tau_\alpha$, 
the system is trapped in a metastable state where
particles just undergo vibrational motion. In case of MC, large and　
small particles are hardly exchanged in this timescale, implying that
$\tau_{\rm ex}\gg \tau_{\alpha}$, thus the effect of the particle
exchange is negligible. The metastable state is then well
described by the FGA. On the contrary, using the SMC with $r$
close enough to unity, large and small particles are easily exchanged, implying
that $\tau_{\rm ex}\ll \tau_{\alpha}$. In this case, the EGA provides a
good description of the metastable state. Within these
assumptions, the efficiency of the SMC is explained by the larger value
of $\varphi_d^{\rm EGA}$ with respect to $\varphi_d^{\rm FGA}$. Note that it is
known that the SMC works only for binary mixtures of sufficiently small
size ratio $r\approx 1.2$~\cite{PhysRevE.63.045102} or continuous
polydisperse system~\cite{andrea2017model}. For binary mixtures of large
size ratio $r\approx 1.4$, the SMC gives a compatible result with that
of the MC~\cite{ozawa2018configurational}.  This cannot be explained by
our phase diagram, Fig.~\ref{133823_19Oct18}, where the difference
between $\varphi_d^{\rm EGA}$ and $\varphi_d^{\rm FGA}$ does not vanish
even around $r\approx 1.4$. The same result is also obtained by
MCT~\cite{szamel2018theory}. The inefficiency of SMC at large $r$ should instead
be attributed to the fact that the assumption $\tau_{\rm ex}<\tau_{\alpha}$ does not hold for
larger $r$, even for the SMC, because the exchange moves are never accepted.
The EGA thus no longer gives a good description of the metastable state.
Unfortunately, because $\tau_{\rm ex}$ is a purely dynamical quantity, it
cannot be calculated by a static theory such as the RLT. It would be
interesting to extend dynamical theories of the glass
transition, such as the mode-coupling theory (MCT), to calculate
$\tau_{\rm ex}$ and reconcile this discrepancy between theoretical and
numerical results.


\ack We thank M.~Ozawa and G.~Szamel for interesting discussions. This
project has received funding from the European Research Council (ERC)
under the European Union's Horizon 2020 research and innovation
programme (grant agreement n.723955-GlassUniversality).  

\appendix

\section{Verlet-Weis approximation for binary mixtures}
\label{130420_28Nov18}

Here we review the binary version of the Verlet-Weis (VW) approximation.  
We label here the two species by indices $i,j\in\{1,2\}$.
As for a
 one-component system, we assume that $g_{ij}(r)$ can be
written as
\begin{eqnarray}
& g_{ij}^{\rm VW}(r) = \theta(r-\sigma_{ij})\left[g^{\rm PY}(\xi r) + \Delta g_{ij}(r)\right],\new
& \xi = \left(\frac{\varphi}{\varphi^*}\right)^{1/3},\new
& \varphi^* = \varphi- \frac{\varphi^2}{16} \ , \new
& \Delta g_{ij}(r) = \frac{A_{ij}}{r}e^{-b_{ij}(r-\sigma_{ij})}\cos(b_{ij}(r-\sigma_{ij})) \ ,
\end{eqnarray}
where
\begin{eqnarray}
 A_{ij}/\sigma_{ij}= g_{ij}^{\rm CS}(\sigma_{ij})-g_{ij}^{\rm PY}(\xi \sigma_{ij},\varphi^*)\ ,  
\end{eqnarray}
where $g_{ij}^{\rm CS}$ and $g_{ij}^{\rm PY}$ denote the results of the
Carnahan-Starling approximation (CS)~\cite{santos2005equation} and
Percus-Yevick approximation (PY)~\cite{lebowitz1964exact},
respectively. We determine $b_{ij}$ from the consistency of the
compressibility equations:
\begin{eqnarray}
 \pdiff{\beta P^{\rm CS}}{\rho} &= 1-\sum_{ij}x_i x_j\rho\int d\bm{r} c_{ij}^{\rm VW}(\bm{r})= \sum_{ij}\left[ \delta_{ij}- \rho_i c_{ij}^{\rm VW}(k=0)\right]x_j\new
 &= \sum_{ij}\left[\bm{I}+\bm{H}(0)\right]_{ij}^{-1}x_j= \sum_{ij}x_i x_j S_{ij}^{-1}(0) \ . \label{163556_11Nov17}
\end{eqnarray}
where
\begin{eqnarray}
 \left[\bm{I}+\bm{H}(k)\right]_{ij} &= \delta_{ij} + \rho_i h_{ij}(k) \ , \new
 S_{ij}(k) &= \delta_{ij}x_j + \rho x_i x_j h_{ij}(k) \ .
\end{eqnarray}
Here, $S_{ij}(k)$ denotes the structural factor and $h_{ij}(k)$ denotes the
Fourier transform of the pair correlation function. For a 
binary mixture, we have
\begin{eqnarray}
 S_{11}^{-1} &= \frac{1}{D}\left(x_2 + \rho x_2^2 h_{22}\right) \ ,\new
 S_{12}^{-1} &= S_{21}^{-1} = -\frac{1}{D}\rho x_1 x_2 h_{12} \ ,\new
 S_{22}^{-1} &= \frac{1}{D}\left(x_1 + \rho x_1^2 h_{11}\right) \ ,
\end{eqnarray}
where
\begin{eqnarray}
 D &= (x_1 +\rho x_1^2 h_{11})(x_2 + \rho x_2^2 h_{22})-x_1^2 x_2^2 h_{12}.
\end{eqnarray}
At large $\varphi$, the compressibility has a large value, which implies
that
\begin{eqnarray}
 \lim_{k\to 0}D(k)&= \lim_{k\to 0}(x_1 +\rho x_1^2 h_{11}(k))(x_2 + \rho x_2^2 h_{22}(k))-x_1^2 x_2^2 h_{12}(k)
 \approx 0.\label{164718_11Nov17}
\end{eqnarray}
We shall determine the value of $b_{ij}$ from this condition.
First, note that $h_{ij}(k)$ can be decomposed as
\begin{eqnarray}
 h_{ij}(k) &= \int dr e^{ikr}\left[g_{ij}(r)-1\right]= h_{ij}^{\rm PY}(k) + \Delta h_{ij}(k) \ ,\new
 h_{ij}^{\rm PY}(k) &= \int dr e^{ikr} \left[\theta(r-\sigma_{ij}/\xi)g_{ij}^{\rm PY}(\xi r)-1\right] \ ,\new
 \Delta h_{ij}(k)&=
 \int dr e^{ikr} \left[\theta(r-\sigma_{ij})-\theta(r-\sigma_{ij}/\xi)\right]g^{\rm PY}(\xi r) \new
& + \int dr e^{ikr}\theta(r-\sigma_{ij})\Delta g_{ij}(r) \ .
\end{eqnarray}
Substituting the above equations into Eq.~(\ref{164718_11Nov17}), we have 
\begin{eqnarray}
 D(0) &= D^{\rm PY}(0) + O(\Delta h_{ij}(0)).
\end{eqnarray}
Because $D(0)$ and $D^{\rm PY}(0)$ are expected to have small values,
$O(\Delta h_{ij})$ terms should also vanish.  The simplest condition is then
\begin{eqnarray}
 \Delta h_{ij}(0)  &= 0 \ .
\end{eqnarray}
This condition can be solved for $b_{ij}$, leading to
\begin{eqnarray}
 b_{ij} &= \frac{24A_{ij}}{\sigma_{ij}^2\varphi g^{\rm PY}(\sigma_{ij},\varphi^*)}
 \approx \frac{24A_{ij}}{\sigma_{ij}^2\varphi^* g^{\rm PY}(\sigma_{ij},\varphi^*)} \ .
\end{eqnarray}



\section*{References}
\bibliographystyle{iopart-num.bst} \bibliography{./reference}

\end{document}